\documentclass[aip,graphicx]{revtex4-1}
\pdfoutput=1
\usepackage{graphicx}
\usepackage{siunitx}
\usepackage{booktabs}
\usepackage{float}
\restylefloat{table}

\begin{document}


\title{Real time Markov chains: Wind states in anemometric data.} 



\author{P. A. S\'{a}nchez-P\'{e}rez }
\email[]{pesap@ier.unam.mx}
\affiliation{Licenciatura en Ingener\'ia en Energ\'ias Renovables, Universidad Nacional Aut\'onoma de M\'exico}


\author{M. Robles}
\email[]{mrp@ier.unam.mx}
\author{O. A. Jaramillo}
\email[]{ojs@ier.unam.mx}
\affiliation{Instituto de Energ\'{i}as Renovables,  Universidad Nacional Aut\'oinoma de M\'exico \\
	Privada Xochicalco SN 62580, Temixco, Mor., M\'exico.}



\date{\today}

\begin{abstract}

The description of wind phenomena is frequently based on data obtained from anemometers, which usually report the wind speed and direction only in a horizontal plane. Such measurements are commonly used either to develop wind generation farms or to forecast weather conditions in a geographical region. Beyond these standard applications, the information contained in the data may be richer than expected and may lead to a better understanding of the wind dynamics in a geographical area.  In this work we propose a statistical analysis based on the wind velocity vectors, which we propose may be grouped in “wind states” associated to binormal distribution functions. We found that the velocity plane defined by the anemometric velocity data may be used as a phase space, where a finite number of states may be found and sorted using standard clustering methods. The main result is a discretization technique useful to model the wind with Markov chains. We applied such ideas in anemometric data for two different sites in Mexico where the wind resource is considered reliable. The approximated Markov chains of both places give a set of values for transition probabilities and residence times that may be regarded as a signature of the dynamics of the site.

\end{abstract}


\maketitle 

\section{Introduction}
\label{sec1}

It is well known that Markov chains are the simplest mathematical models for random phenomena.
However, at the same time they are a class of stochastic process rich enough to be applied at many different problems and considered the theoretical base of many  other models  \cite{Norris1997}.
A Markov chain can be understood as an stochastic process without memory, therefore the future state of a system only depends on the present situation and not in its history.
These models assume that the system under study can only visit a finite number of states, with probabilities that does not depend on time; then, the system is moving  between states with a given probability.

Markov chain approaches are frequently present in wind speed forecasting.
Modern probabilistic models based on autoregressive (AR) and moving average (MA) models, frequently uses Markov chains in order to estimate some of the multiple parameters needed \cite{Jung2014}.
Markov chains are the base of forecasting methods \cite{Damico2014}, but they are also useful to generate synthetic wind speed time series  needed in simulations approaches \cite{Shamshad2005,Sahin2001,Aksoy2004, Damico2013}. 

In order to apply a Markov chain model to  climatological data, wind speed have to be converted into discrete states.
The most common way to do this is splitting the data associating  wind speed values between two limits for a given state \cite{Shamshad2005,Sahin2001,Damico2013}.
Although this separation is practical and give useful results, those states do not represent any physical property of the system beyond a discrete speed.
Also  it is important to mention that in many studies the analysis focus only in modeling the wind speed, perhaps because is the main quantity needed to forecast the power generation or to estimate energy production.
Some other studies introduce changes in wind direction, using Markov chains and defining some goegrafical channels, which classify of the angles in discrete values \cite{Castino1998}; these channels treated as Markov states allow to calculate probabilities of transition between them. 

The wind direction seems to be a very important parameter for being taken in account in the wind assessment
when the statistical analysis report that the speed frequencies do not follow a standard Weibull distributions, and then one may suspect the influence of many physical wind modes related with different directions \cite{Jaramillo2004}.
In such cases forecasting become difficult and more complex methods are needed\cite{Erdem2011}.

This work is addressed to present and discuss  a strategy to analyze multi-modal wind data and generate physical states that can be used in Markov chain models of wind.
We will base the discussion on two data sets containing one year of anemometric data at two different locations in Mexico.
The structure of this work is as follow: first we will present the study cases to introduce the need of consider the  velocity vector time series and its statistical properties. 
This will lead us to establish the definition of ``wind states'' and to propose the use of clustering methods to classify them.
Then we will show the results of building Markov Chains for the data and illustrate the procedure to get the simplest stochastic model.
A further discussions and conclusions will be addressed at the end of the work.

\begin{figure}
	\includegraphics[height=4.1cm]{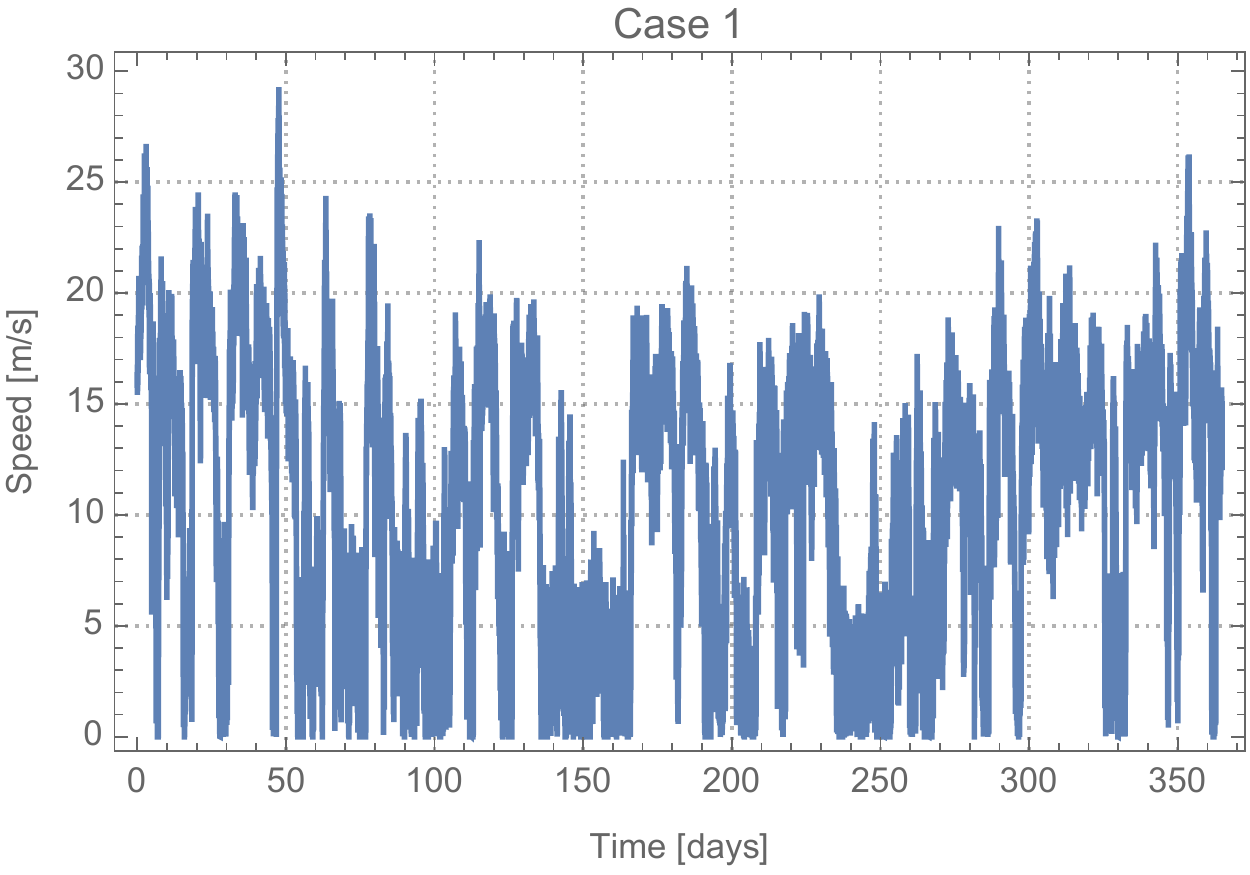}
	\includegraphics[height=4.1cm]{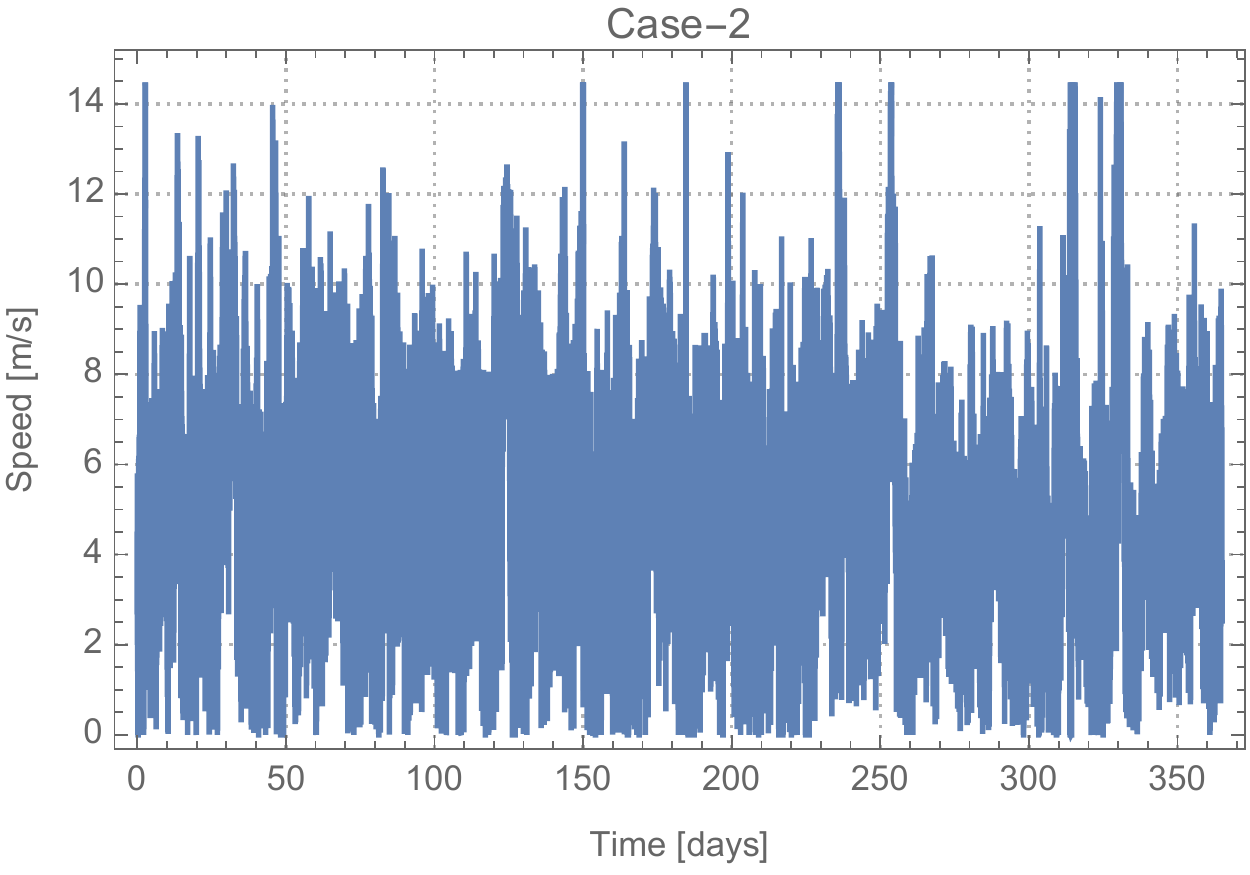}\\
	\includegraphics[height=4cm]{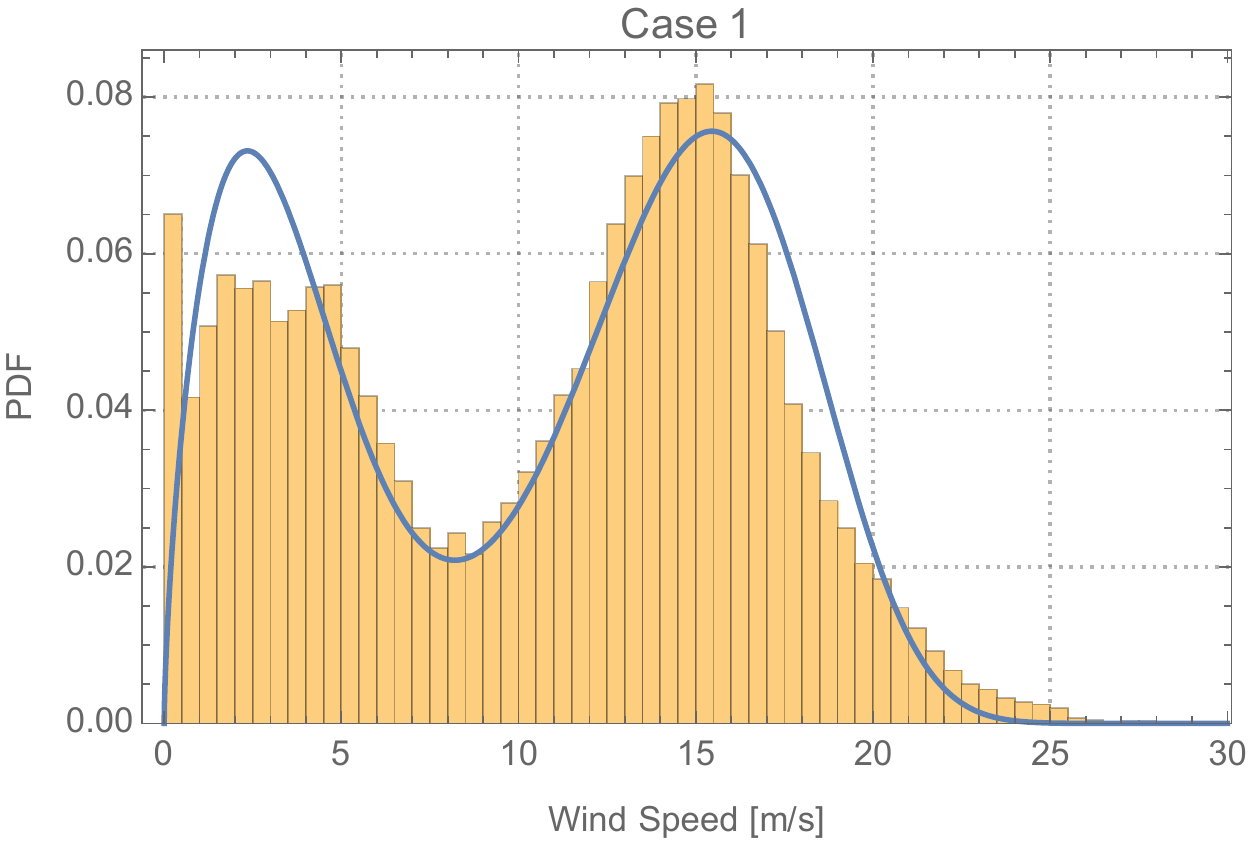}
	\includegraphics[height=4cm]{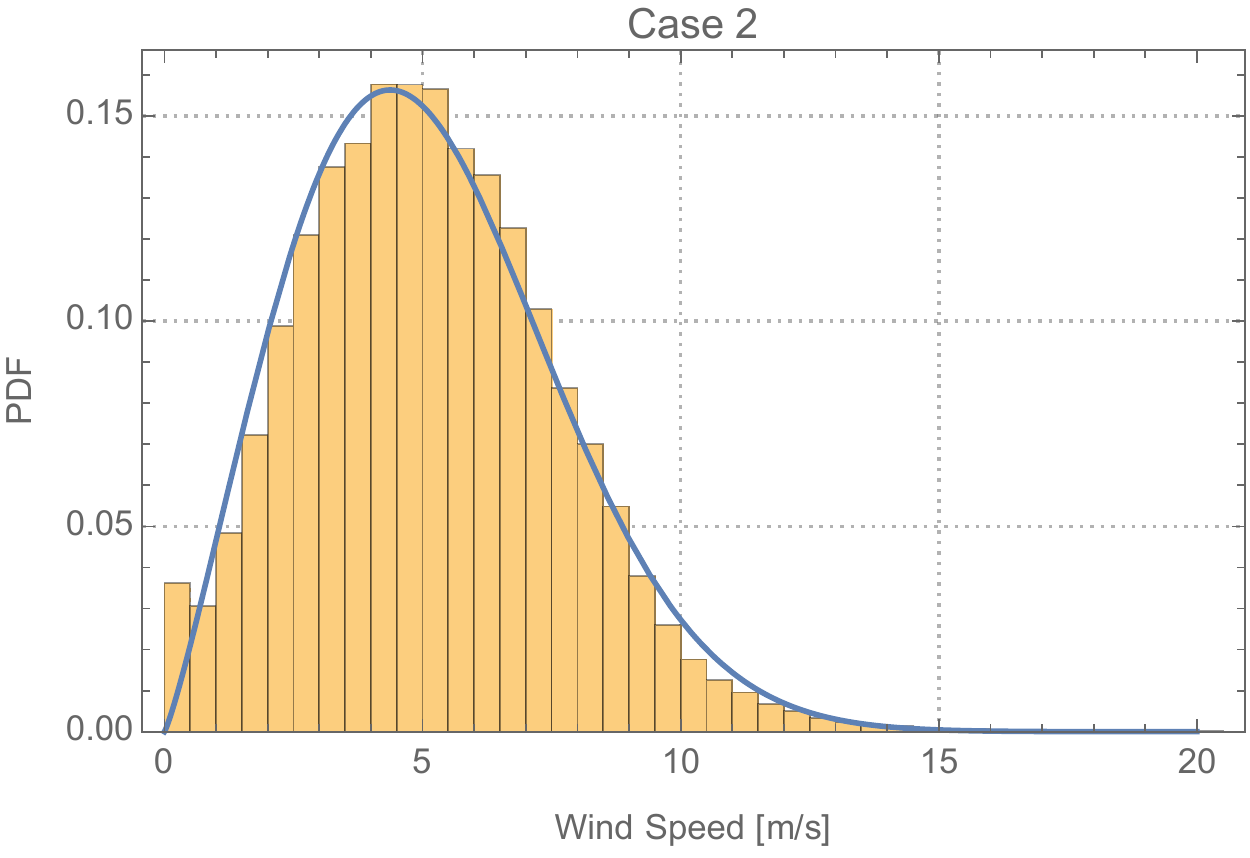}
	\caption{Probability distribution functions for the two time series chosen as case of study. The histograms where computed directly from the data set and the continuous lines are the best fit functions taken obtained in \cite{Jaramillo2004,Jaramillo2004a}.}
	\label{hist2d}
\end{figure}

\section{Definition for ``wind states".}

We choose two study cases. In the first one we use wind data analyzed by Jaramillo and Borja in \cite{Jaramillo2004}.
These data correspond to one year of measurement, with \SI{10}{\minute} averaged measure at the wind farms located in la Ventosa, Oaxaca at \SI{32}{\meter} above the ground level.
Case 2 consist in one year of measurement with the same time resolution in a wind station at Baja California, Mexico \cite{Jaramillo2004a} at \SI{10}{\meter} above the ground level.
These two cases were chosen because their different behavior on his statistical properties.
Case 1 was reported to behave as a multi-modal function in which wind speed distribution can be approximated as a superposition of two Weibull distributions.
This behavior is associated to the changes on wind direction.
In contrast, case 2 is a mono-modal and wind speed probability distribution function (PDF), fits with a single Weibull distribution.
The PDFs of both data sets are shown in Figure \ref{hist2d}. 
For case 1 fit parameters for a bi-Weibull distribution can be found in reference \cite{Jaramillo2004} and for case 2 in reference \cite{Jaramillo2004a}.

\begin{figure}
	\begin{center}
		\includegraphics[height=5cm]{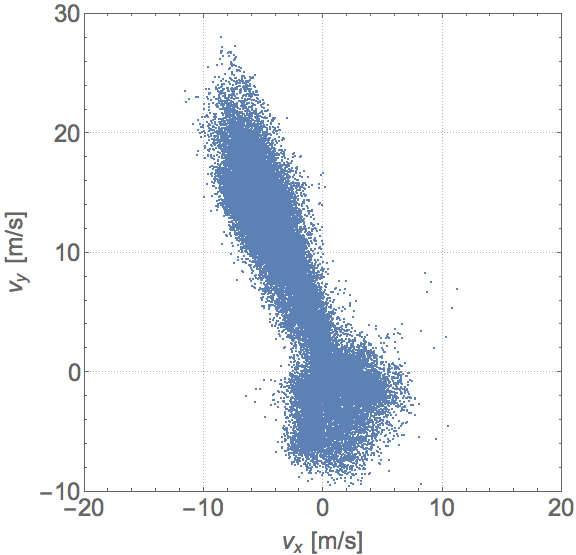}
	\end{center}
	\caption{Scatter plot for the velocity vectors for  one year wind speed data on case 1.}
	\label{cumul1}
\end{figure}

Let's start with  case 1 and instead of using a wind rose, we will analyze the velocity components with a scatter plot of velocity vectors.
Using the wind speed and direction, usually given in weather data, one can construct an average vectorial velocity for each measurement, by simply doing $\vec{V_i}=v_i(\cos \theta_i, \sin \theta_i)$, where $v_i$ and $\theta_i$ are the speed and direction data in the time step $i$. 
This is a new collection from two time series, one per each velocity and his associated direction.
Now, if we plot all points in the plane $v_x$-$v_y$ instead of visualizing the time series separately, we would obtain the plot shown in Figure \ref{cumul1}.

It is imporant to point out that figure~\ref{cumul1} resembles a phase space in the statistical mechanics sense.
The wind velocity has vast regions that can not be visited ever, and the points accumulate around well defined regions that may be regarded as ``states''.
If we recall that the angle from the data set is measured from the north direction, the first direct observation of this plot is that a preferred  direction exist.
In other words, there is a preferred ``state'' in the NNW direction and this state is where the higher speeds were found. The same result was found previously in \cite{Jaramillo2004}.

\begin{figure}
	\begin{center}
		\includegraphics[width=6.0cm]{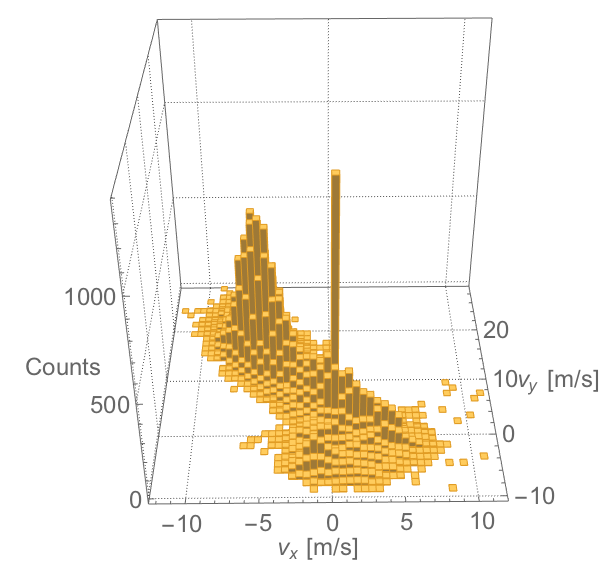}
		\includegraphics[width=6.0cm]{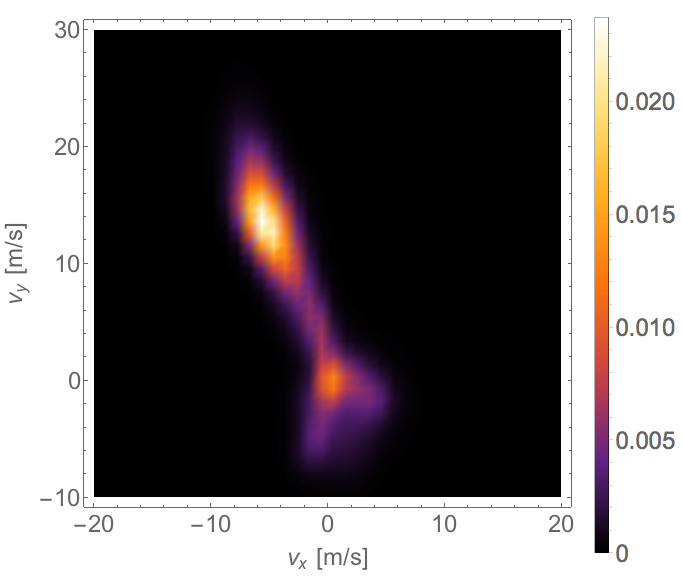}\\
		(a)\hspace{6cm}(b)
	\end{center}
	\caption{Three dimensional frecuenccy histogram (a) and smooth density plot (b),   for the velocity vectors in the phase plane, for the study case 1.}
	\label{hist3d1}
\end{figure}

To drive the analysis deeper we can plot a three dimensional histogram with his smooth density  shown in Figure~\ref{hist3d1}(a) and (b) respectively.
Figure \ref{hist3d1}(a) counts the number of velocity vectors that lie in a specif squared bin, Figure~\ref{hist3d1}(b) is a colored numerical interpolation for the PDF.
If the two dimensional velocity space where an ``ensemble'' of all possible wind ``micro-states'', we could conclude that for long time evolution, there are stable ``macro-states'' represented by regions where the system is more probably to be found. 
In this case, it is possible to say that, there are two main states, the first located around the origin, which means the absence of wind or where the system is in repose. 
The second state appears in the third quadrant (the NNW direction) and the wind speeds are comprehended in the interval from 2 to \SI{25}{\meter\per\second} always in the same direction.
These two main states could not be the only ones, there could be more ``states'' with lower probabilities that may be also relevant for specific purposes.
In order to look after these states, we propose a numerical strategy that we will discuss in section \ref{2:Clustering}.

\begin{figure}
	\begin{center}
		\includegraphics[width=5.0cm]{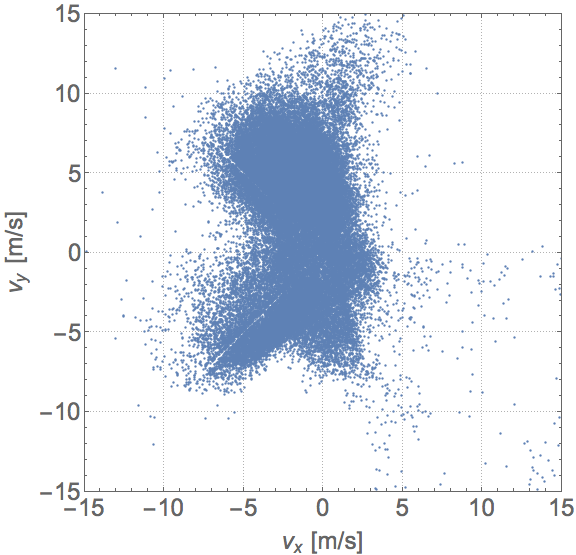}	
		\includegraphics[width=5.8cm]{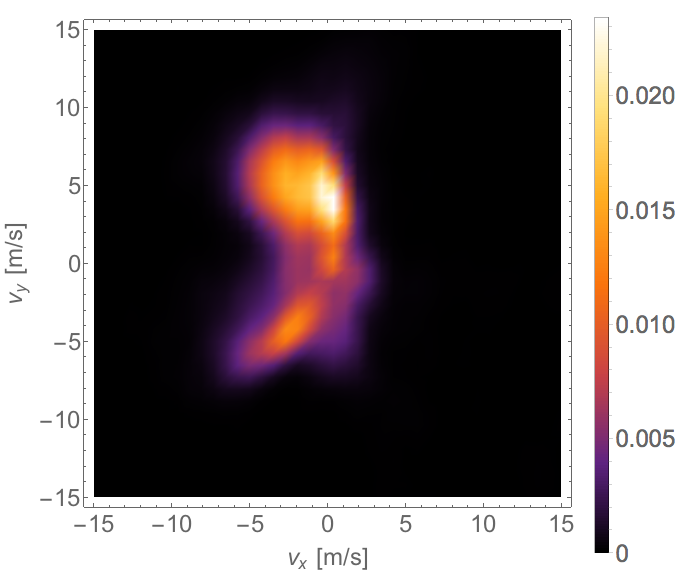}\\
		(a)\hspace{5cm}(b)\\
		\includegraphics[width=7.0cm]{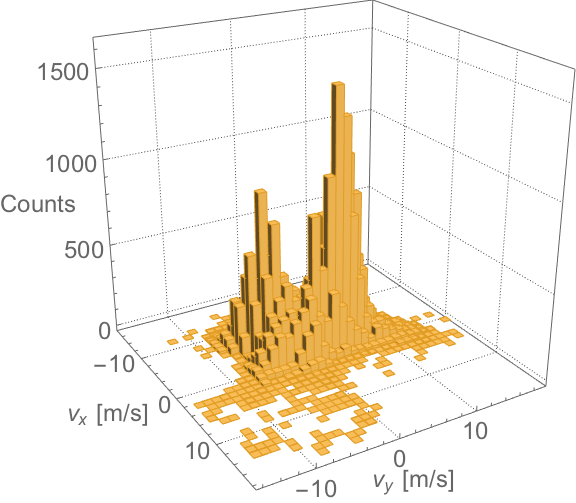}\\
		(c)
	\end{center}
	\caption{Scatter plot (a), smooth densiyt plot (b) and the three dimensional histogram (c) for the velocity vectors in phase plane, for the study case 2.}
	\label{hist3d2}
\end{figure}

The same analysis can be applied to case 2 and it give the results in Figure \ref{hist3d2} (a), (b) and (c).
First, clearly there are more than one marcro-states compatible with the figures. 
It is possible to tell at a glance that at least four relevant cumulus of points can be located.
As in the first case, the repose state is also present and probably would be in any case. 
Obviously, the worst situation for power generation is when the repose state is visited more than any other state.
On the opposite side, the best situation for power generation could be close to case 1, because one dominant state exist with good speeds in a narrow direction.
Therefore, case 2 is in the middle, is a place with three main states that can generate some wind power, the speed on each is moderated and is located in three different orientations; it is possible to say that an specific wind generator could be designed to take advantage of this situation.

Before to conclude this section, based on the last qualitative analysis, we will introduce a more precisely definition of what we understand as a wind state.
It should be referred to a region in the $v_x$-$v_y$ plane where the wind repeatedly arrives. 
Physically they are not the same as a thermodynamic equilibrium state or an out of equilibrium steady states. 
They have an internal dynamics that changes all the time, but they are a property of the long time series and are determined by the geographical and climatic conditions in the site.
A wind state is a region in the velocity phase space that contain the accessible wind velocities having a common probability distribution function that characterize them as a cluster.
The PDFs for all the states are like a landscape where the velocity vectors move.
The transition dynamics between different wind states is highly stochastic and probably depends on the seasonal cycles or many other climate changes. 
The physics behind is certainly very complex and therefore beyond the scope of this work. 
Nevertheless, it is possible to quantify the previous analysis and try to establish a link with stochastic models. 
The first step is try classify the wind data in the states without know much about the nature of them. 
To do so, in the next section we use a clustering method, which have become and option to classify time series data efficiently \cite{rani2012, Clifton2012, Lund1999} .

\section{Clustering wind states from wind speed data}
\label{2:Clustering}

The only information to establish the landscape of the wind velocity phase space are the meteorological data of the site. 
In previous section the qualitative analysis let us know that the velocity points tend to get together in cumulus.
Therefore a direct way to try to sort them in the different wind states is applying a clustering algorithm. 
Those methods are usual in different computational science problems including, data mining, machine learning, pattern recognition or bio-informatics. 

\begin{figure}
	\begin{center}
		\includegraphics[width=5.5cm]{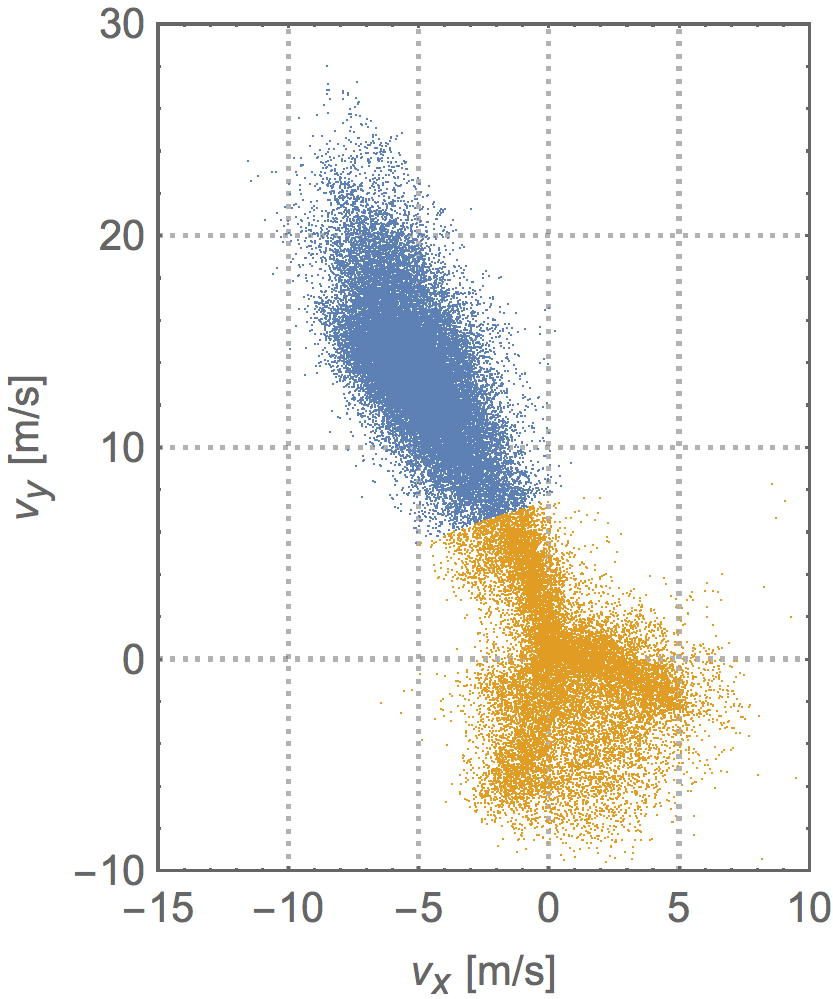}
		\includegraphics[width=5.5cm]{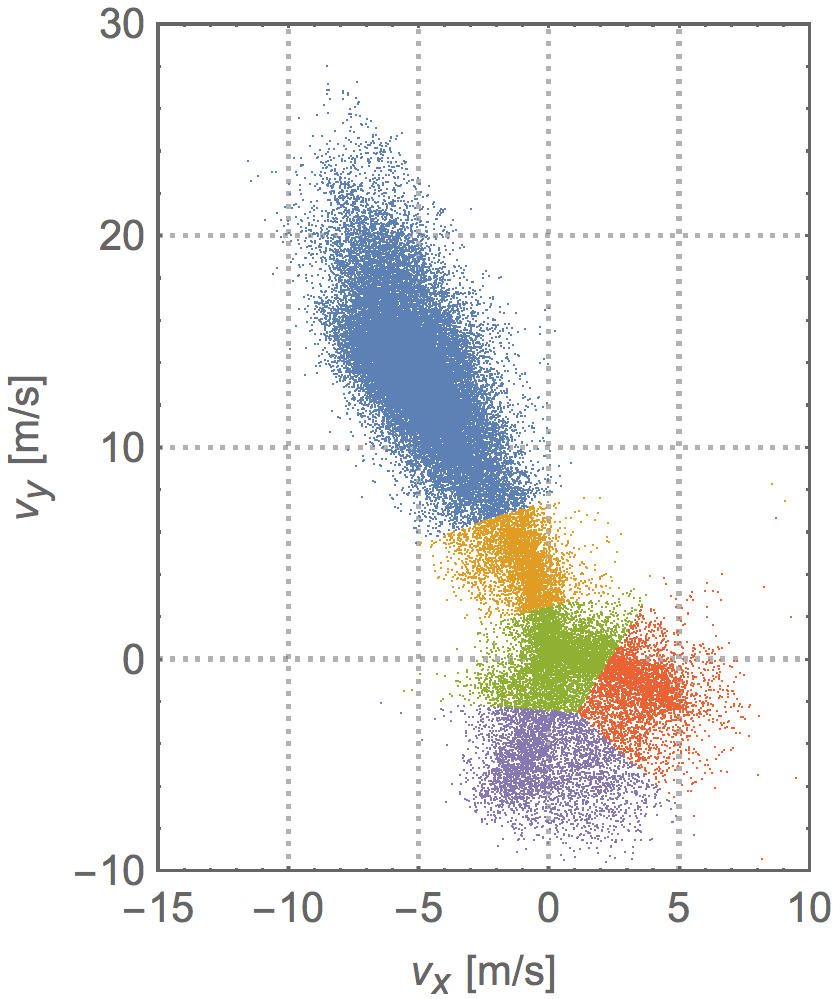}\\
		(a)\hspace{4cm}(b)
	\end{center}
	\caption{Two steps $k$-means clustering applied to case 1 data. In the first step (a) the algorithm was applied to the whole data requesting two partitions. For the second step (b) the algorithm was applied only to the data in the lower partition using $k=4$.}
	\label{kmeans-1}
\end{figure}

\begin{figure}
	\begin{center}
		\includegraphics[height=4.8cm]{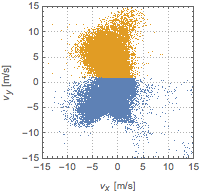}
		\includegraphics[height=4.8cm]{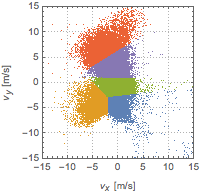}\\
		(a)\hspace{4.5cm}(b)
	\end{center}
	\caption{Two steps $k$-means clustering applied to case 2 data. In the first step (a) the algorithm was applied to the whole data requesting two partitions. For the second step (b) the algorithm was applied again to the two new partitions requesting $k=3$ an $k=2$ respectively.}
	\label{kmeans-2}
\end{figure}

Currently, there are several effort to apply clustering methods to time series and wind data such the works reported in the literature \cite{rani2012, Clifton2012, Lund1999, Piazza2010}, one of the most used is the  $k$-means clustering algorithm \cite{MacQueen1967}.
The method is used to partition a set of $n$-points in $k$-clusters, classifying with a criterion of the nearest mean. 
Although it is a difficult problem, modern heuristic algorithms can find rapidly an optimized solution. 
To apply the $k$-means method, is necessary propose the optimum number of clusters present in the set ($k$), since this is a whole different problem in computational science, we decided to use as a first approximation the higher count of peaks in the histograms obtained in previous section.
It worth to mention that the idea of applying $k$-means algorithm is not new, can be found previously for example in \cite{Leite2006}, to reduce the number of speed states in the time series. 
For case 1, the smooth density histogram of the Figure \ref{hist3d1}(b) indicates the presence of two main states.
This will help us to choose the number of cluster for $k$-means algorithm.
Applying  using $k=2$ one may easily find the initial partition shown in Figure \ref{kmeans-1}(a).
To discover the internal structure close to the repose state, a second application of $k$-means using $k=4$, but only for this partition, we obtain Figure \ref{kmeans-1}(b). 
It is important to notice that $k$-means clustering do not give an unique solution, so we need to be careful with the partition.
If initially we request $k=5$ to the entire time series we will obtain a different configuration of clustering.
Depending on the centroid initialization clustering will vary because the algorithm can converge to a local minimum that may not be representative.
The last partition from this procedure reveals a structure closer to the expected. 
However, this is a limitation to the method because it is needed to see internal structure of the time series before clustering.
A similar procedure was applied for case 2. 
A graphical resume is depicted in Figure \ref{kmeans-2}. 
A final partition  having five states seems to be the compatible with the explored histograms.

It is important to note that each partition obtained from the three dimensional histogram can be associated with one wind state. Therefore the clustering procedure illustrated may be assumed as a classification method.
Once we classified the states we can build a continuous Markov Chain using the time series of the wind states as discussed in the next section.
To do this, one more assumption has to be done: the points in a given state distribute with Gaussian bivariate function.
Also, it is possible to find the local PDF that can be assumed as a Gaussian bivariate function by simplicity in order to correlate the wind speed and direction.

A Gaussian multivariate function for a vectorial space $\vec{r}$,  with dimension $d$, is defined as:
\begin{equation}
f(\vec{r})=\frac{1}{\sqrt{(2\pi)^d |\Sigma|}} \exp\left( \frac{1}{2} (\vec{r}-\vec{\mu})^T \Sigma ^{-1} (\vec{r}-\vec{\mu}) \right),
\label{eq1} 
\end{equation}
where $\vec{\mu}$ is the centroid of the distribution and $\Sigma$ the covariance matrix, whose components follow the expected value:
\begin{equation}
\Sigma_{i,j}=E[(r_i-\mu_i)(x_j-\mu_i)].
\end{equation}
The super index $T$ indicates the transpose of the vector and $\Sigma^{-1}$ is the inverse of the  matrix $\Sigma$. 
For two dimensional case, where $\mu=(\mu_x,\mu_y)$ and the $\Sigma$ matrix is
\[ \Sigma = \left( \begin{array}{cc}
\sigma_{xx} & \sigma_{xy}  \\
\sigma_{yx} & \sigma_{yy} 
\end{array} \right).\] 

Lets apply the following statistical procedure to each partition of the data in case 1:
\begin{enumerate}
	\item Compute the centroid and covariance matrix.
	\item Calculate the Gaussian bi-variate function for each partition.
	\item Compose the total distribution function as a superposition of all previous Gaussian, re-normalizing with their respective weights in proportion to the total number of points.
\end{enumerate} 

\begin{figure}
	\begin{center}
		\includegraphics[width=8cm]{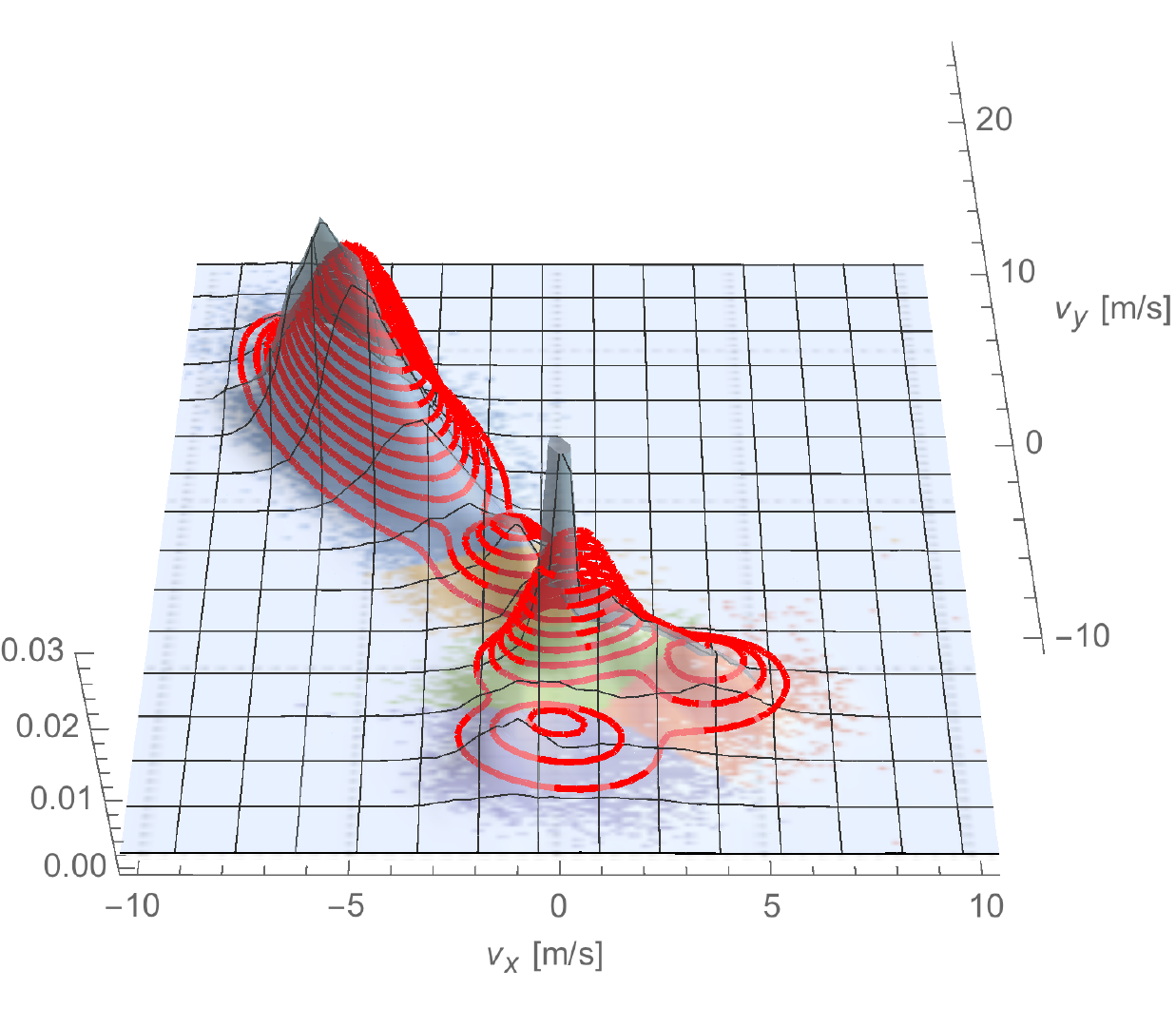}
	\end{center}
	\caption{Probability distribution function obtained from the superposition of Gaussian bi-variate functions (contour lines in red) compared with the numerical distribution obtained directly from real data (light blue surface) in case 1. As a guide the velocity points were included in the background $v_x$-$v_y$ plane. }
	\label{states1}
\end{figure}

\begin{figure}
	\begin{center}
		\includegraphics[width=8cm]{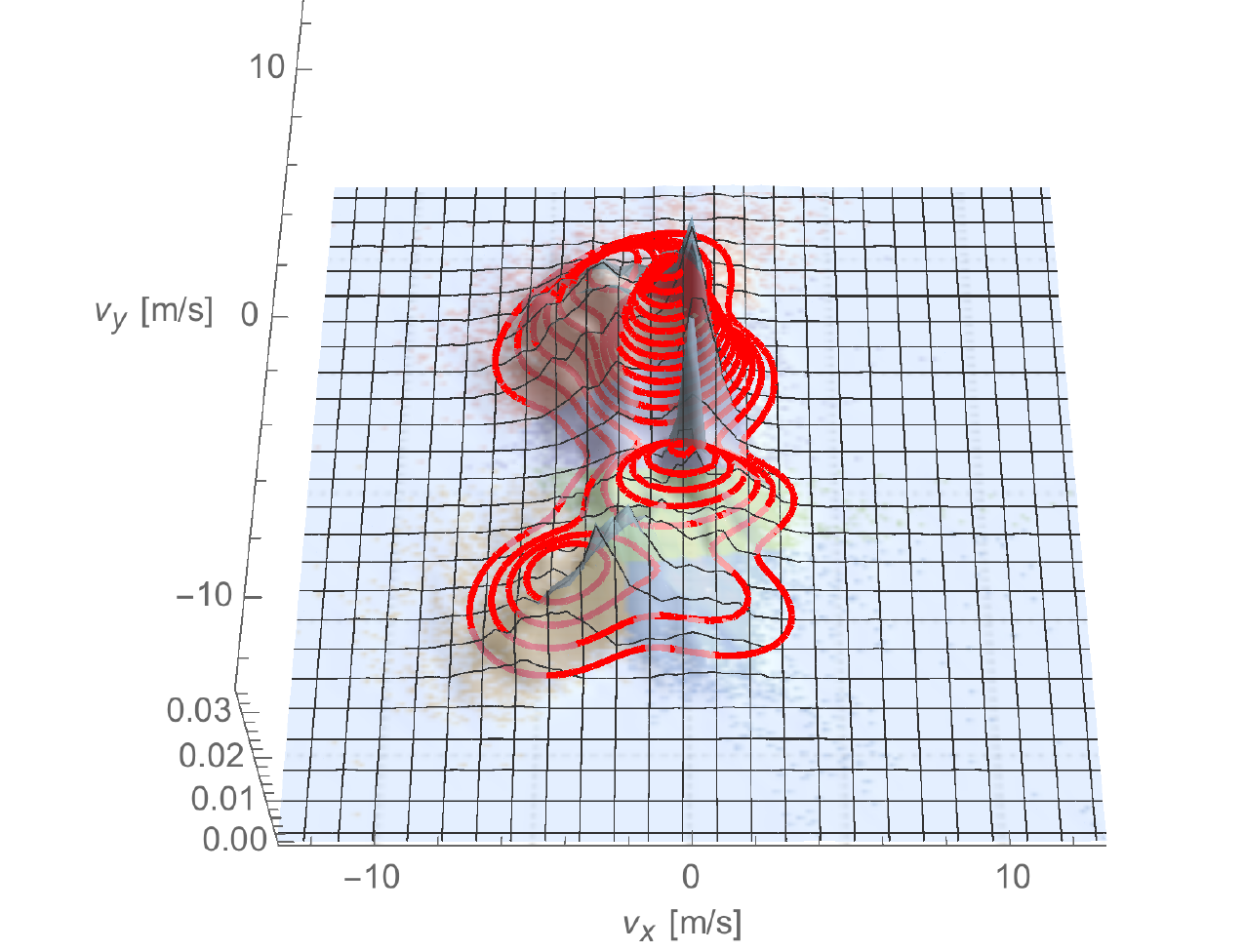}
	\end{center}
	\caption{Probability distribution function obtained from the superposition of Gaussian bi-variate functions (contour lines in red) compared with the numerical distribution obtained directly from real data (light blue surface) of case 2. As a guide the velocity points were included in the background $v_x$-$v_y$ plane. }
	\label{states2}
\end{figure}

The Figure \ref{states1} contain the total probability distribution function obtained for case 1 data,  compared with a direct numerical computation from the real data. 
It is important to point out that the Gaussian bi-variate functions can be applied without considering a fitting method.
Repeating the same steps on case 2 data it is possible to depict the Figure \ref{states2}.  
These data have a more complex structure and do not agree as well compared with the case 1.  
However, applying this methodology to  case 2 gives some information about the landscape; 
The numerical values of $\vec{\mu}$ and the matrix $\Sigma$ for both cases can be found in the appendix.

Beyond the precision of the method, there are two concepts that worth to remark.
Firstly, the notion of state must be associated with a PDF and this could be described, at least on a first approximation, by considering Gaussian distributions. 
In some way resembles the state functions in quantum mechanics and statistical mechanics, but yet without a master equation known. 
The second concept is more computational, the use of clustering methods, adapted to this problem, together with the availability of wide historical data bases, may substitute the lack of dynamical equations in this complex problem, allowing to build up efficient numerical methods reliable for practical applications.   

To conclude this section we can say that clustering methods may be one key to study the wind states and determine their state functions. 
The advantage of use them is that can be easily connected with data mining algorithm to apply it systematically to all geographical areas where data is available. 
In the next section we will introduce the dynamics between states in both study cases when we interpret the dynamics as Markov chains.

\section{Markov chains for the wind states.}
After the clustering process, the velocity points are sorted in one and only one partition set. 
In other words, they are associated with one wind state. 
Each wind state can now be labeled arbitrary, then we decided to use the numbers given in the column ``states" in tables \ref{parm1} and \ref{parm2}, as can be seen in Appendix \ref{ap1}.

\begin{figure}
	\begin{center}
		\includegraphics[width=12.6cm]{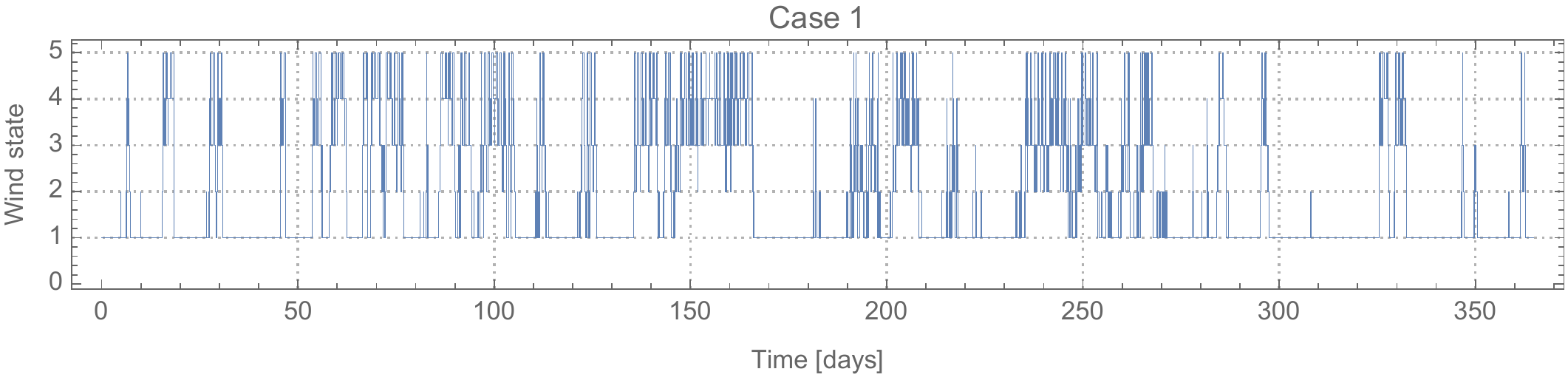}\\
		(a)\\
		\includegraphics[width=12.6cm]{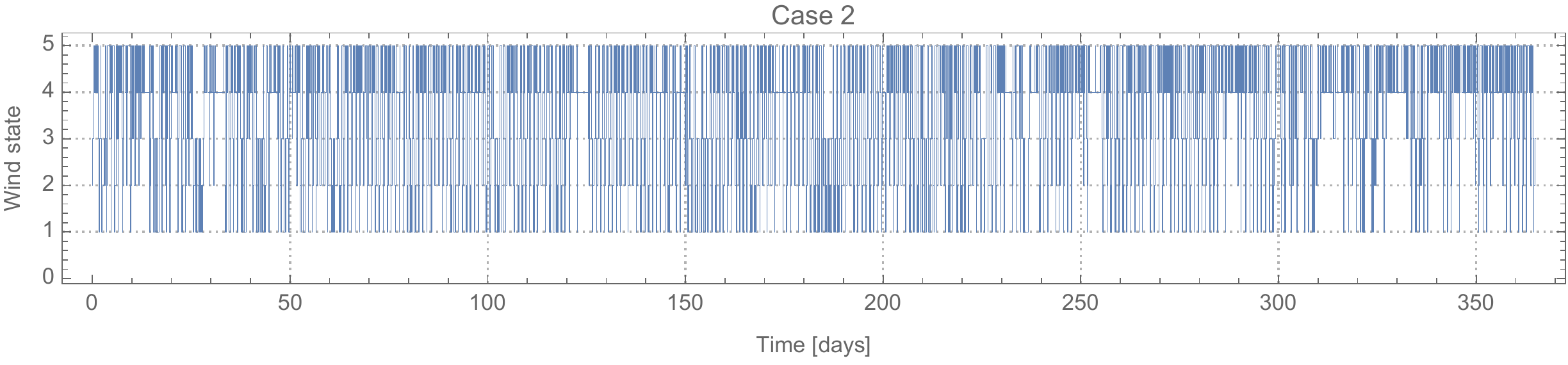}\\
		(b)\\
	\end{center}
	\caption{Time Markov chains associated with both case of study. Case 1(a) and case 2 (b). }
	\label{Mkch}
\end{figure}

A real time Markov chain is developed from the original wind data time series. 
The chain is assembled by considering the association of each vector with  its state to which it belongs. 
The Figure \ref{Mkch} presents the real time Markov chains for both study cases. 
The first information coming from figures, at glance, is that case 2 is a lot more complex than the case 1 since it looks more unstable.
For the case 1 the state 1 is preferred over the rest. 
To quantify this appreciations we computed the transition probability matrix between the states for both cases, assuming that the sequence as a discrete Markov chain. 
The results are reported in Tables \ref{tm1} and \ref{tm2}. 
In Table \ref{tm1} it is important to point out the zero entries correspond for transitions between state 1 to 3 and 5, from state 4 to 1 and from 5 to 1.
The existence of these unreachable or forbidden transitions must be considered because they can have direct technological applications in the design of wind generators and control algorithms.

To go forth, using the real Markov chains we can measure the total time $t_r$ and the mean life time $t_m$ for the wind in each state. 
A summary of these values are presented in Table \ref{times}, where it is important to highlight the $232.417$ days that the wind in case 1 stays in the state 1, together with a mean lifetime close to 0.9 days of the time series.
This state is close to ideal to produce wind power, is a state with speeds around  \SI{14}{\meter\per\second} with directions in a narrow sector visited many days in a year and each time is reached it stays in average almost one day. 
It looks possible to locate a wind turbine without yaw system and therefore does not have a significant reduction on the electric generation instead of a wind turbine with yaw system included.  
In contrast none of the states, in  case 2, stays in a state for more than $2.4$ hours, in average although that some of them reach $122$ days in total. 
This imply that a generator should be always tracking the different states, which may imply a higher maintenance investment, due to the increase of costs in the yaw operation system. Clearly, the information contained in the Markov chain is relevant for technology purposes.

The knowledge of mean life time for each state allows to compute the q-matrix that generates the chains, if were assumed as continuous Markov processes.  
The off diagonal elements of the q-matrix, represent the time rate at which the systems leave the state $i$ and arrive to state $j$, $q_{ij}$. 
The diagonal elements $q_{ii}$ are the negative average rate at which the system leaves the state $i$, therefore if the mean life time of states $i$ is $t_{mi}$ then $q_{ii}=-1/t_{mi}$ \cite{Norris1997}. 
The computation then is simple, for a state $i$ find the transition rate to other state $j$ and then multiply by the rate of leaving state $i$, $-q_{ii}$. 
For case 1 and 2 the resulting q-matrices are reported in tables \ref{qm1} and \ref{qm2}, also in the Appendix \ref{ap1}.  

For case 1, the state 1, which would be the best candidate for energy production has at most $1.15$ transitions per day, the rest do up to $19$ transitions. 
In case 2 is not possible to find a state with less that $10$ transitions per day, which can be regarded as a disadvantage. 

Beyond the last interpretations, the $Q$-matrices contain the information needed to create a synthetic Markov chain with close physical properties to the real system. 
It is a the simplest model and may not produce an accurate forecast, however, it seems a powerful quantity to characterize the states of wind in a geographical situation and could be the base to create machine learning algorithms to automatic search good sites for wind energy production. 

\begin{figure}
	\begin{center}
		\includegraphics[width=4.8cm]{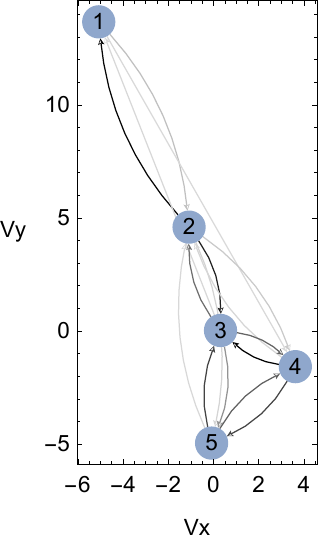} 
		\includegraphics[width=4.8cm]{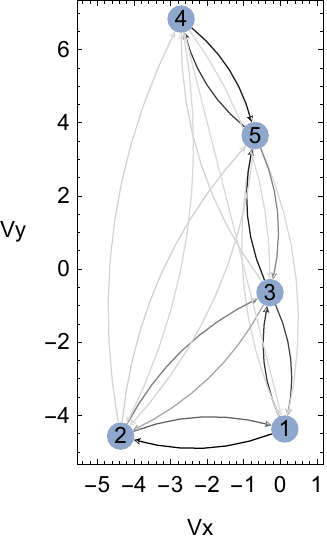}\\
		(a) \hspace{4.5cm}(b)\\
	\end{center}
	\caption{Graphs associated with the $Q$-matrices for  case 1(a) and case 2 (b). }
	\label{grphs}
\end{figure}

The $Q$-matrices can be represented as graphs. 
The normal procedure to build them is placing the states in nodes or vertexes linked by arrows that indicate the transitions allowed. 
With this idea in mind we represented the q-matrices obtained in a special form. 
First we placed the vertexes of the graph, representing the states, right  in the plane of velocities, just in the center of the bi-normal distribution which represent the associated state function. 
The links between states are drown as gray arrows using a scale associated with the value of the $q_{ij}$ entry in the q-matrix; 
black color was set to the maximum of all the entries, excluding the diagonals, and the minimum to a gray value that yet allow to distinguish their presence in the graph. 
The results are figures \ref{grphs}(a) and (b). These graphs are a visual description of the mechanics of wind in the analyzed sites.

For the case 1, the graph \ref{grphs}(a) immediately address the issue on whether is really useful a tracking method for an optimal generator. 
Graphically we realize that state 1 is the best in terms of wind speed, and the changes in direction when a transition takes place  from it, are always to states with lower velocity. 
Moreover, the fluxes indicate that there is a pressure to return either to state 1 or 2. In contrast case 2 (figure \ref{grphs}(b)) is more stochastic. 
There are at least three states with good speed potential (states 1, 4 and 5) but, all of them point to different directions and have important fluxes to the other states. 
This interpretation may change the evaluation of the wind resource done by traditional methods.

To finish this section we must remark the fact that, although  the wind states seems to exists independently of the time, the time series for wind velocities always show seasonal effects, for that reason  Markov chains are not the best representation. However, the transition and $q$-matrices are a numerical properties that may represent a signature of the mechanics of wind that may be useful beyond the physical understanding. 

\section{Discussion}

At this point we reach the question on how this results may contribute either to the understanding of wind phenomena or to improve traditional wind assessment. Clearly, the definition of  wind state given here is only based on the assumption that wind velocities are described by  two dimensional Gaussian modes that may be obtained from the cluster partition. It is important to point out that the physical causes for each state are indeed a complex problem beyond the scope of this work. However, it is possible to propose how to characterize physically the states and try to find out how useful are to estimate the power generation. If the state $i$ has a Gaussian component $f_i(v_x,v_y)$ as given in equation (\ref{eq1}), the wind power per unit area for the state may be calculated by:
\begin{equation}
\frac{P_i}{A}=\frac{1}{2} \rho \int_{0}^{\infty}\int_{0}^{\infty} |\vec{v}|^3 f_i(v_x,v_y) dv_x dv_y,
\end{equation}
where $|\vec{v}|=(v_x^2+v_y^2)^{1/2}$ and $\rho$ is the air density that in average may be considered as $1.21\ kg/m^3$, at sea level. 

If $m$ is the number of wind states the total power of the site is simply $P=\sum_{i=1}^{m} P_i$ and the relative power of a state may be adopted as $p^*_i=P_i/P$. This relative power give us the information about how the state may weight in the total energy production and then, it can be used as a figure of merit.

\begin{figure}
	\begin{center}
		\includegraphics[width=7.3cm]{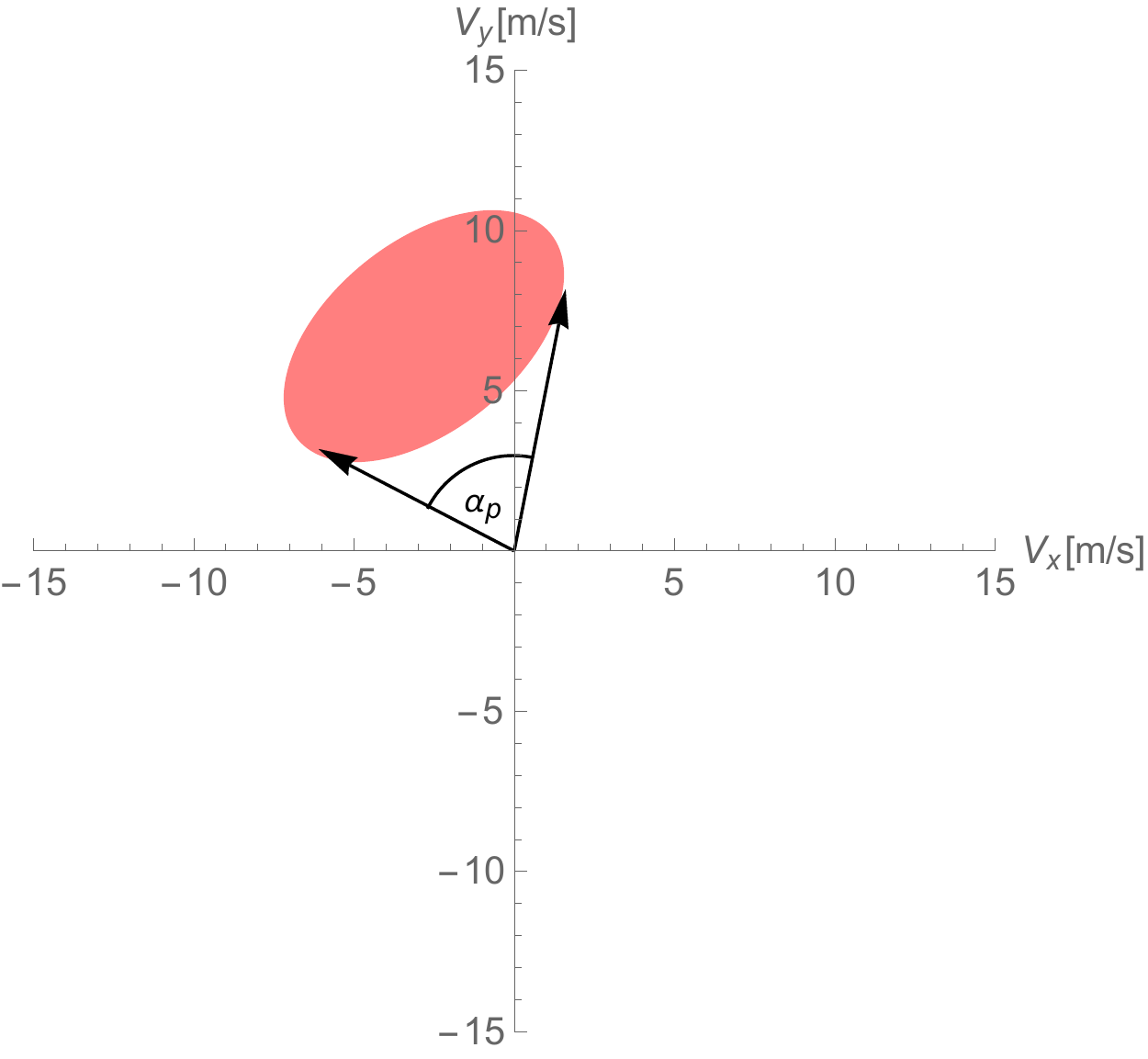} 
	\end{center}
	\caption{Graphical construction of $\alpha_p$ in the state 4 for the study case 2.}
	\label{apert}
\end{figure}

\begin{figure}[b]
	\begin{center}
		\includegraphics[width=6.5cm]{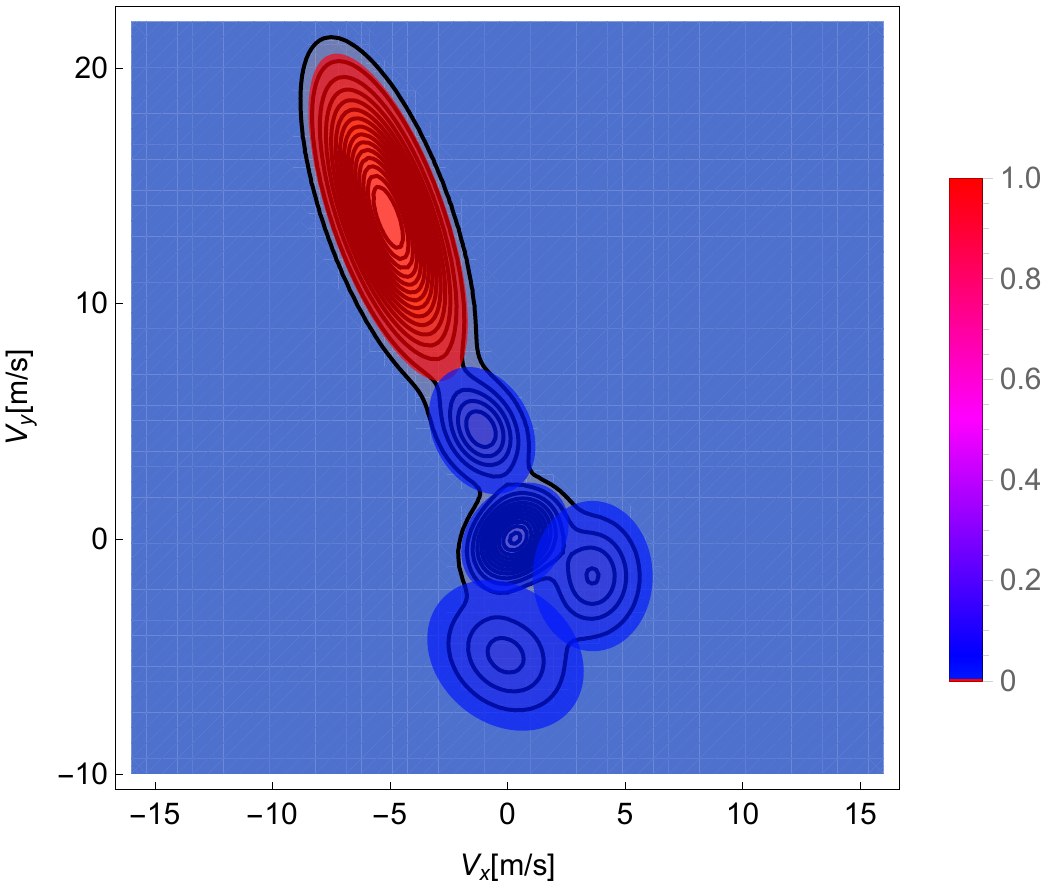} 
		\includegraphics[width=6.5cm]{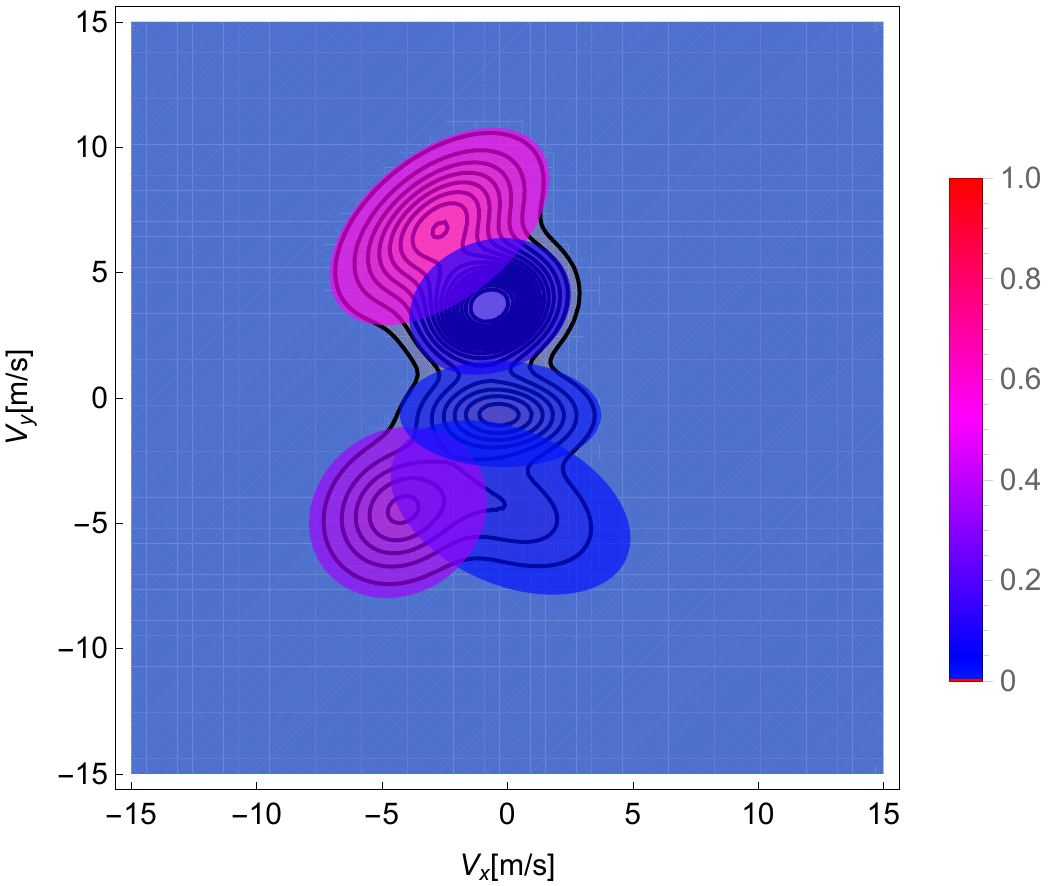}\\
		(a) \hspace{4.8cm}(b)\\
	\end{center}
	\caption{Elliptical regions where Gaussian components of each state covers $90 \%$ of the data associated with it for the study case 1 (a) and 2 (2). Ellipses where colored using relative power $p^*_i$ with the color code shown in legends.}
	\label{ellips}
\end{figure}

Another important effect is the angular dispersion of  wind states. An ideal state is such that concentrate the total power in a very narrow angular sector. We can define the ``aperture" ($\alpha_p$) of a state as the angle subtended by the ellipse formed with the isoline of the $i-$esime Gaussian component that covers the $90\%$ of data in the state, and represent the angular region where the wind blows with more probability, as is shown in figure \ref{apert}. Note that this new definition may enhance the information given by the standard methods, such wind rose graphs, where the frequency of wind directions are grouped in discrete circular sectors, with the same angle, and colored with the wind speed. This statement may be observed in Figure \ref{ellips}, where the ellipses used for the compute $\alpha_p$ are presented, for both study cases, colored with the relative power. These graphs show clearly the regions where wind produces most of the energy and therefore, together with the dynamics shown in Figure \ref{grphs}, could assist, for example in the optimization of tracking methods, defining the regions where generator could not produce more energy than the consumed by yaw system, and then could be completely avoided, balancing the tracking algorithms to maximize production. 

The values of $\alpha_p$ obtained for both study cases are in Table \ref{merit}, with the relative power $p^*_i$ and the direction $\theta$ of center of Gaussian component. All the angles are measured in degrees and $\theta$ values are relative to north direction ($v_y$). 
We can remark some observations: the numbers immediately say that state 1 on the study case 1 is the best of all, because it maximizes the relative power and minimizes the aperture. That was already known, but now this conclusion may be achieved trough an automatic quantitative procedure easy to implement. Note that, since on both cases the state 3 defines regions that contains the origin, the aperture must be fixed on $360^\circ$, \textit{i.e.} the maximum possible value. In contrast the $p^*_3$ is very low, therefore this state maximize aperture and minimize power, and it is the worst situation for energy generation. That was also known, but again the procedure may detect such states. Summarizing, the Gaussian components assumed as state functions may allow to define quantitative numerical strategies to analyze the energy production in a specific location.

\begin{table}
	\caption{Relative power, aperture and average direction for all the states in study cases 1 and 2.}
	\begin{tabular}{|c|ccc|ccc|}
		\hline
		{} & \multicolumn{3}{c}{Case 1} & \multicolumn{3}{c}{Case 2} \vline \\ \hline
		State& $p^*_i \ [\%]$ &$\alpha_p \  [^\circ]$& $\theta \ [^\circ]$& $p^*_i \ [\%]$ &$\alpha_p \  [^\circ]$ & $\theta\ [^\circ]$	\\	
		1 & 98.68 & 21.95 & 20.37 &1.92 & 119.90 & 181.70 \\
		2 & 0.43 & 58.88 & 13.22 & 34.75 & 64.81 & 136.30 \\
		3 & 0.03 & 360.00 & 274.24 & 0.85 & 360.00 & 156.70 \\
		4 & 0.29 & 93.25 & 246.30 & 54.16 & 73.60 & 21.64 \\
		5 & 0.57 & 82.74 & 0.97 & 8.33 & 103.10 & 10.63 \\
		\hline
	\end{tabular}
	\label{merit}
\end{table}

\section{Conclusions}

We have presented a numerical analysis of wind velocity vectors obtained from anemometric time series. The analysis, based on the $k-$means clustering method was applied to two study cases. The data, accumulated for one year, sort in clusters well delimited in the 2D velocity space, and may be associated with bivariate Gaussian distributions. These relations allow us to introduce the concept of ``wind state". The geographical and meteorological characteristics of the site should determine the formation of those states in a complex manner, however, their characteristics may be approximated through the computational analysis shown. It is reasonable to think that the bigger the wind data base is in the site, more detailed states can be found, therefore the wind states should be time independent in analogy with thermodynamic states, hence the method may be applied the same for different time-stepping discretization  instead of $10\ \min$ averaged measurements.  The local dynamics may be described in a rough manner by the transition probabilities  between them. The instate dynamics is subject of further investigation.
It seems that the wind states introduce a ``natural" discretization of the velocity that may convert the time series in discrete Chains, that in first approximation can be assumed as Markovian process. With this assumption the transition matrices and Q-matrices can be obtained for any set of data, regardless the time resolution or how long the history of data is, and represent in a simple way a first approximation to their stochastic dynamic behavior. In certain way these matrices are a footprint of the dynamics in a given site, and it is possible to be unique, like a signature. 
We are convinced that the method illustrated here is an alternative analysis to the traditional evaluation methods, and provides new information about the wind. Therefore, it can be useful to understand the properties of wind, 
 to improve the wind resource assessment and drive new control methods  in order to enhance energy production. Finally, it is worth to pointing out that physical properties added to the wind states, such relative power and aperture angle,  allow to compare different wind regions, and to choose the most convenience for energy production.

\section*{Acknowledgment}

We would like to acknowledge Dr. Rafael Campos and Dr. Mishael S\'anchez-P\'erez for the useful discussions and suggestions. Also we acknowledge Dr. Maximiliano Valdez Gonz\'alez for the computational assistance. 

\appendix
\section{Tables.}\label{ap1}

In this appendix we include the tables with the numerical calculation associated with the normal bivariate state functions, transition matrix, characteristic times and q matrices for the two study cases presented in the text. All the values were computed from the time series, sorting the points in the state through the k-means algorithm described. 
\subsection{Normal bivariate distribution parameters for each state.}
\begin{table}[H]
	\caption{Gaussian parameters for wind states for the study case 1.}
	\begin{tabular}{ccccccc}
		\toprule
		State& $\mu_x [m/s]$ &$\mu_y[m/s]$& $\Sigma_{xx}[(m/s)^2]$& $\Sigma_{xy}[(m/s)^2]$& $\Sigma_{yx}[(m/s)^2]$ & $\Sigma_{yy}[(m/s)^2]$	\\	
		\midrule
		1&	-5.0746 & 13.6675 & 2.4915 & -3.3270 & -3.3270 & 10.5284 \\
		2&	-1.0804 & 4.5985 & 1.1161 & -0.3334 & -0.3334 & 1.5873 \\
		3&	0.3163 & 0.0234 & 1.1334 & 0.2858 & 0.2858 & 1.2095 \\
		4&	3.6258 & -1.5895 & 1.4108 & -0.0093 & -0.0093 & 2.2184 \\
		5&	-0.0843 & -4.9619 & 2.4047 & -0.5016 & -0.5016 & 2.2301 \\
		\bottomrule
	\end{tabular}
	
	\label{parm1}
\end{table}

    \begin{table}[H] 
    	\begin{tabular}{ccccccc}
    		\toprule
    		State& $\mu_x [m/s]$ &$\mu_y[m/s]$& $\Sigma_{xx}[(m/s)^2]$& $\Sigma_{xy}[(m/s)^2]$& $\Sigma_{yx}[(m/s)^2]$ & $\Sigma_{yy}[(m/s)^2]$ \\
    		\midrule
    		
    		
    		1&   0.1269 & -4.3762 & 4.9710 & -1.2755 & -1.2755 & 2.6258 \\
    		2&   -4.3630 & -4.5714 & 2.7266 & 0.3326 & 0.3326 & 2.5311 \\
    		3&   -0.2794 & -0.6473 & 3.5139 & -0.0438 & -0.0438 & 0.9712 \\
    		4&   -2.7106 & 6.8333 & 4.1543 & 1.8232 & 1.8232 & 3.3550 \\
    		5&   -0.6847 & 3.6476 & 2.2533 & 0.3284 & 0.3284 & 1.6114 \\

    		\bottomrule
    	\end{tabular}
    	\caption{Gaussian parameters for wind states for the study case 2.}
    	\label{parm2}
    \end{table}
    
    \subsection{Transition matrix between states.}
    
    \begin{table}[h]
    	\centering
    	\begin{tabular}{cccccc}
    		\toprule
    		State & 1 & 2 & 3 & 4 & 5 \\	
    		\midrule
    		
    		1&       0.9921 & 0.0079 & 0.0000 & 0.0001 & 0.0000 \\
    		2&      0.0680 & 0.8665 & 0.0590 & 0.0049 & 0.0015 \\
    		3&       0.0001 & 0.0341 & 0.9056 & 0.0376 & 0.0226 \\
    		4&       0.0000 & 0.0010 & 0.0701 & 0.8815 & 0.0473 \\
    		5&       0.0000 & 0.0002 & 0.0485 & 0.0386 & 0.9127 \\

    		\bottomrule
    	\end{tabular}
    	\caption{Transition matrix for the real time Markov chain obtained from the case 1 data.}

    	\label{tm1}
    \end{table}

    \begin{table}[h]
    	\centering
    	\begin{tabular}{cccccc}
    		\toprule
    		State& 1 & 2  & 3 & 4 & 5  \\
    		\midrule
    		
    		1&        0.8600 & 0.0794 & 0.0600 & 0.0000 & 0.0006 \\
    		2&       0.0446 & 0.9158 & 0.0338 & 0.0021 & 0.0037 \\
    		3&        0.0549 & 0.0191 & 0.8548 & 0.0021 & 0.0690 \\
    		4&       0.0001 & 0.0008 & 0.0004 & 0.9287 & 0.0701 \\
    		5&        0.0002 & 0.0016 & 0.0306 & 0.0569 & 0.9107 \\
    		
    		\bottomrule
    	\end{tabular}
    	\caption{Transition matrix for the real time Markov chain obtained from the case 2 data.}
    	
    	\label{tm2}
    \end{table}
    \pagebreak
    \subsection{Characteristic times.}
    
    \begin{table}[H]
    	\centering
    	\begin{tabular}{ccccc}
    		\toprule
    		States	      &\multicolumn{2}{c}{Case 1} &\multicolumn{2}{c}{Case 2}\\
    		\midrule
    		& $t_{r} $ (days) & $t_{m}$ (days) & $t_{r}$(days) & $t_{m} $(days) \\ 
    		1    & 232.417 & 0.868973  &  37.3542  & 0.0496071 \\
    		2    & 26.9444 & 0.0520163 &  50.7361  & 0.0825203 \\
    		3    & 50.9583 & 0.0735329 &  53.3403  & 0.0477641 \\
    		4    & 26.5486 & 0.0586062 &  100.861  & 0.0973563 \\
    		5    & 28.0694 & 0.0795168 &  122.701  & 0.0777575\\
    		\bottomrule
    	\end{tabular}
    	\caption{Total residence time $t_r$ and mean life time $t_m$ from the Markov chains for the two study cases. }
    	\label{times}
    \end{table}
    
    \subsection{Q-Matrices.}
    
    \begin{table}[h]
    	\centering
    	\begin{tabular}{cccccc}
    		\toprule
    		State & 1 & 2 & 3 & 4 & 5 \\	
    		\midrule
    		
    		1&    -1.1508 & 1.1421 & 0.0000 & 0.0087 & 0.0000 \\
    		2&   9.7979 & -19.2247 & 8.4990 & 0.7052 & 0.2227 \\
    		3&  0.0196 & 4.9060 & -13.5993 & 5.4162 & 3.2576 \\
    		4&    0.0000 & 0.1507 & 10.0947 & -17.0630 & 6.8177 \\
    		5&    0.0000 & 0.0356 & 6.9827 & 5.5576 & -12.5760 \\
    		
    		\bottomrule
    	\end{tabular}
    	\caption{Q-matrix for the real time Markov chain obtained from the case 1 data.}
    	
    	\label{qm1}
    \end{table}
    
    \begin{table}[H]
    	\centering
    	\begin{tabular}{cccccc}
    		\toprule
    		State & 1 & 2 & 3 & 4 & 5 \\	
    		\midrule
    		1&          -20.1584 & 11.4311 & 8.6470 & 0.0000 & 0.0803 \\
    		2&          6.4236 & -12.1182 & 4.8670 & 0.2956 & 0.5320 \\
    		3&          7.9238 & 2.7602 & -20.9362 & 0.3004 & 9.9518 \\
    		4&          0.0099 & 0.1091 & 0.0595 & -10.2716 & 10.0931 \\
    		5&          0.0326 & 0.2363 & 4.4009 & 8.1906 & -12.8605 \\

    		\bottomrule
    	\end{tabular}
    	\caption{Q-matrix for the real time Markov chain obtained from the case 2 data.}
    	\label{qm2}
    \end{table}



%
%

%


%

\end{document}